\documentclass[aps,prd,twocolumn,groupedaddress,showpacs,preprintnumbers,draft]
{revtex4}
\usepackage{epsf}
\begin{document}
\def\Box{\nabla^2}
\def\ie{{\em i.e.\/}}
\def\eg{{\em e.g.\/}}
\def\etc{{\em etc.\/}}
\def\etal{{\em et al.\/}}
\def\S{{\mathcal S}}
\def\I{{\mathcal I}}
\def\mL{{\mathcal L}}
\def\H{{\mathcal H}}
\def\M{{\mathcal M}}
\def\N{{\mathcal N}}
\def\O{{\mathcal O}}
\def\cP{{\mathcal P}}
\def\R{{\mathcal R}}
\def\K{{\mathcal K}}
\def\W{{\mathcal W}}
\def\mM{{\mathcal M}}
\def\mJ{{\mathcal J}}
\def\mP{{\mathbf P}}
\def\mT{{\mathbf T}}
\def\mR{{\mathbf R}}
\def\mS{{\mathbf S}}
\def\mX{{\mathbf X}}
\def\mZ{{\mathbf Z}}
\def\eff{{\mathrm{eff}}}
\def\Newton{{\mathrm{Newton}}}
\def\bulk{{\mathrm{bulk}}}
\def\brane{{\mathrm{brane}}}
\def\matter{{\mathrm{matter}}}
\def\tr{{\mathrm{tr}}}
\def\nr{{\mathrm{normal}}}
\def\implies{\Rightarrow}
\def\half{{1\over2}}
\newcommand{\da}{\dot{a}}
\newcommand{\db}{\dot{b}}
\newcommand{\dn}{\dot{n}}
\newcommand{\dda}{\ddot{a}}
\newcommand{\ddb}{\ddot{b}}
\newcommand{\ddn}{\ddot{n}}
\newcommand{\ba}{\begin{array}}
\newcommand{\ea}{\end{array}}
\def\be{\begin{equation}}
\def\ee{\end{equation}}
\def\bea{\begin{eqnarray}}
\def\eea{\end{eqnarray}}
\def\bs{\begin{subequations}}
\def\es{\end{subequations}}
\def\g{\gamma}
\def\G{\Gamma}
\def\vp{\varphi}
\def\mpl{M_{\rm P}}
\def\ms{M_{\rm s}}
\def\ls{\ell_{\rm s}}
\def\lp{\ell_{\rm pl}}
\def\l{\lambda}
\def\gs{g_{\rm s}}
\def\d{\partial}
\def\co{{\cal O}}
\def\sp{\;\;\;,\;\;\;}
\def\spa{\;\;\;}
\def\r{\rho}
\def\dr{\dot r}
\def\dt{\dot\varphi}
\def\e{\epsilon}
\def\k{\kappa}
\def\m{\mu}
\def\n{\nu}
\def\om{\omega}
\def\tn{\tilde \nu}
\def\p{\phi}
\def\vp{\varphi}
\def\P{\Phi}
\def\r{\rho}
\def\s{\sigma}
\def\t{\tau}
\def\x{\chi}
\def\z{\zeta}
\def\a{\alpha}
\def\b{\beta}
\def\de{\delta}
\def\bra#1{\left\langle #1\right|}
\def\ket#1{\left| #1\right\rangle}
\newcommand{\stt}{\small\tt}
\renewcommand{\theequation}{\arabic{section}.\arabic{equation}}
\newcommand{\eq}[1]{equation~(\ref{#1})}
\newcommand{\eqs}[2]{equations~(\ref{#1}) and~(\ref{#2})}
\newcommand{\eqto}[2]{equations~(\ref{#1}) to~(\ref{#2})}
\newcommand{\fig}[1]{Fig.~(\ref{#1})}
\newcommand{\figs}[2]{Figs.~(\ref{#1}) and~(\ref{#2})}
\newcommand{\GeV}{\mbox{GeV}}
\def\ricci{R_{\m\n} R^{\m\n}}
\def\riemann{R_{\m\n\l\s} R^{\m\n\l\s}}
\def\triemann{\tilde R_{\m\n\l\s} \tilde R^{\m\n\l\s}}
\def\tricci{\tilde R_{\m\n} \tilde R^{\m\n}}
\title{Canonical constraints on leptonic CP violation using UHCR neutrino fluxes}
\author{K.R.S. Balaji$^{1,2,3}$ \email[Email:]{balaji@hep.physics.mcgill.ca},
Gilles Couture$^2$\email[Email:]{couture.gilles@uqam.ca},
Cherif Hamzaoui$^2$ \email[Email:]{hamzaoui.cherif@uqam.ca}}
\affiliation{ $^1$ Department of Physics, McGill University, Montr\'eal, QC,
Canada H3A 2T8}
\affiliation { $^2$ D\'epartement des Sciences de la Terre et de
l'Atmosph\`ere, Universit\'e du Qu\'ebec \`a Montr\'eal,
C.P. 8888, succ. centre-ville, Montr\'eal, QC, Canada H3C 3P8}
\affiliation {$^3$Physique des Particules, Universit\'e
de Montr\'eal,C.P. 6128, succ. centre-ville, Montr\'eal, QC,
Canada H3C 3J7}
\begin{abstract}
It is shown that one can in principle constrain the CP-violating
parameter $\delta$ from measurements of four independent
$|V_{ij}|^2$'s, or three $|V_{ij}|^2$ and a ratio of two of them,
in the leptonic sector. To quantify our approach, using unitarity,
we derive simple expressions in terms of four independent
$|V_{ij}|^2$'s for $\cos\delta$, and an expression for
$\sin^2\delta$ from $J^2$. Thus, depending on the values for
$|V_{ij}|$ and their accuracy, we can set meaningful limits on
$|\delta|$. To illustrate numerically, if $|V_{\mu 1}|^2$ is close
to 0.1 with a 10\% precision, and if $|V_{e3}|^2$ is larger than
0.005 and for values of $|V_{e2}|^2$ and $|V_{\mu 3}|^2$ that stay
within $\pm 0.1$ of the current experimental data leads to a
bound: $\pi/2\leq|\delta|\leq\pi$. Alternatively, a certain
combination of parameters with values of $|V_{e3}|^2$ larger than
0.01 leads to a closed bound of $73\leq|\delta|\leq 103$. In
general, we find that it is better to use $|V_{\mu 1}|^2$ or
$|V_{\tau 1}|^2$ as the fourth independent $|V_{ij}|^2$ and that
over most of the parameter space, $\delta$ is least sensitive to
$|V_{e3}|^2$. With just three independent measurements (solar,
atmospheric and reactor), it is impossible to set limits on the CP
phase. In this respect, we study the use of ultra high energy
cosmic ray neutrino fluxes as the additional fourth
information. We find that within the SM, neutrino fluxes of all
three flavours will be very similar but that pushing current
neutrino data to their extreme values still allowed, ratios of
cosmic neutrino fluxes can differ by up to 20$\%$; such large
discrepancies could imply negligibly small CP-violation. We also
study a non radiative neutrino decay model and find that the
neutrino fluxes can differ by a factor of up to 3 within this
model and that an accuracy of 10$\%$ on the neutrino fluxes is
sufficient to set interesting limits on $\delta$.
\end{abstract}
\maketitle

\section{Introduction}
One of the significant achievements of particle physics has been our better
understanding of the leptonic flavor mixings. Following some very detailed and
pains taking measurements the neutrino anomaly has now been confirmed;
both in the solar and atmospheric sector \cite{sk}. An immeadiate conclusion is
the existence of new physics  beyond standard model and to introduce a small
neutrino mass. Such an extension leads to the notion of neutrino oscillation
\cite{ponty}
similar in spirit to quark sector. A genuine three flavor analysis, allows
for a solution space consisting of two mass-squared differences
and three angles.
The solution amounts to finding the allowed parameter space for the
solar mixing angle, and the atmospheric mixing angle along
with a strongly constrained reactor angle.
Phenonemologically, best fit values and allowed range for these three mixing
parameters
dictates a pattern that is almost being
negligible (reactor) to moderate (solar) to
maximal (atmospheric). For a review on the analysis, we refer to \cite{nuph}.

Similar to the quark sector, one can have CP-violation in the leptonic sector
due to massive neutrinos. Unlike the quark sector however, the CP violation
in the leptonic sector can come due to both Dirac and Majorana phases in the
mixing matrix, depending on the nature of the neutrinos
\cite{schechter-valle}
Clearly, assessing
the Dirac or Majorana nature of the neutrinos will be an important goal in the
future. In this paper, we consider only CP-violation {\it \`a la Dirac} and do
not address at all Majorana phases. As a pertinent question, how to measure
CP-violation in neutrino oscillations has attracted a lot of  attention in the
past  \cite{sato,lindner} and no doubt,
the search for leptonic CP-violation will be one of the main goals of
experimental particle physics in the years to come.
Given the strong reactor constraints \cite{chooz} it is very
difficult for current neutrino oscillation experiments
to measure CP-violation. Nonetheless,
as a proposal, a measurement is possible by one searching for differences
between neutrino and anti-neutrino survival in a long-baseline experiment and
measuring the spectrum \cite{marciano}.

In the present analysis, we want to explore the possibility of extracting
information on CP violation in the leptonic sector without performing
a direct CP violation experiment. Clearly, such an experiment
will have to be done eventually but in the near future, it might be easier to
measure individual lepton mixing matrix elements ($V_{ij}$). Thus, it
becomes interesting to see how information about these can be translated into
information about CP violation in the leptonic sector and what level of
precision will be required on the $|V_{ij}|^2$'s to set interesting
constraints on $\delta$.

In the next section, we will contrast two parametrizations that represent
leptonic mixing; the usual parametrization in terms of mixing angles
and phases \cite{vpmns}
and another one constructed purely as moduli elements,
where CP violation is not explicit but still present. Following  this,
we will derive expressions that allow one to extract information on
CP violation from measurements of four $|V_{ij}|$'s or combinations of them.
These relations are used to calculate the precision on
the measurements of the $|V_{ij}|^2$'s required in order to set constraints on
$\delta$. We will then consider the use of UHCR neutrino fluxes to
obtain the fourth $|V_{ij}|^2$ needed to set constraints
on $\delta$ \cite{ffoot}.
We also derive general relations about their fluxes and
estimate the precision required in order to set interesting limits on
$\delta$.  To conclude, we will also consider a specific model that digresses
from the SM and will strongly affect the neutrino fluxes that reach the earth
and see how useful this can be in our extraction of the phase $\delta$.

\section{Linking four $|V_{ij}|^2$ and $\delta$}
\subsection{Rephasing invariant parametrization}
We assume that we have three neutrino flavours and consider only
CP-violation {\it \`a la Dirac}. We also assume unitarity in the
mixing matrix even though some popular models of neutrino mass
generation lead to violation of unitarity.

The elements of the lepton mixing matrix are defined as
\begin{eqnarray}
\pmatrix{ \nu_e \cr
\nu_{\mu} \cr
\nu_{\tau} } = \pmatrix{V_{e1} & V_{e2} & V_{e3} \cr
V_{\mu 1} & V_{\mu 2} & V_{\mu 3} \cr
V_{\tau 1} & V_{\tau 2}  & V_{\tau 3} \cr}
\pmatrix{ \nu_1 \cr
\nu_2 \cr
\nu_3 }~.
\label{umix}
\end{eqnarray}
It has been shown before \cite{cherif} that all the information
about mixing and CP violation in the quark sector can be
parametrized in terms of four independent moduli of (\ref{umix}).
These moduli are rephasing invariant and basis independent. A set
of four independent parameters is not unique and any set of four
parameters where we don't have three on the same row or column is
acceptable. Thus, there are nine such ensembles which are allowed.
For the present analysis, our choice is $|V_{e2}|$, $|V_{e3}|$,
$|V_{\mu1}|$ and $|V_{\mu 3}|$. This choice seems appropriate
since we have experimental information on three of the four
parameters.

The exact expressions between the moduli of the elements of the mixing matrix
and the mixing angles are obtained to be:
\begin{eqnarray}
|V_{e3}|^2 & =& sin^2\theta_{13} ~,\\
|V_{e2}|^2 & =& sin^2\theta_{12}~cos^2\theta_{13}~,\\
|V_{\mu 3}|^2 & =&sin^2\theta_{23}~cos^2\theta_{13}~.
\end{eqnarray}

Our current knowledge on the magnitude of the elements of the leptonic mixing
matrix comes from experiments on neutrino oscillation and can be summarized as
follows \cite{nuph}:

\begin{eqnarray}
\sin^2 \theta_{12} & = & 0.3\pm0.08 \nonumber\\
\sin^2 \theta_{23} & = & 0.5\pm0.18 \nonumber\\
\sin^2 \theta_{13} & \leq & 0.05~~{\rm at}~3\sigma
\label{expnos}
\end{eqnarray}
\noindent

\subsection{ Unitarity and $cos(\delta)$}
At the present time, there is no experimental information about $|V_{\mu1}|$
and one can only limit its modulus through unitarity. In the notation of (\ref{umix}), we can write
\begin{eqnarray}
V_{\mu 1} & = &-\frac{1}{1-|V_{e3}|^2}(|V_{e2}||V_{\tau 3}| +
                    |V_{e1}||V_{\mu 3}||V_{e3}|e^{i\delta}) ~,  \nonumber\\
|V_{e1}|& = &\sqrt{1 - |V_{e2}|^2 - |V_{e3}|^2} ~,\nonumber\\
|V_{\tau 3}|& = &\sqrt{1 - |V_{\mu 3}|^2 - |V_{e3}|^2}~.
\label{umix1}
\end{eqnarray}
Expanding up to order $|V_{e3}|^2$, we calculate $|V_{\mu 1}|^2$ to be
\bea
|V_{\mu 1}|^2 &= & |V_{e2}|^2(1-|V_{\mu 3}|^2) +
        |V_{e3}|^2(|V_{e2}|^2 - 3|V_{e2}|^2|V_{\mu 3}|^2 \nonumber\\
        &+&|V_{\mu 3}|^2)+ 2 |V_{e3}||V_{e2}||V_{\mu 3}| \nonumber\\
        &\times& \cos(\delta)\sqrt{1-|V_{\mu 3}|^2}
\sqrt{1-|V_{e2}|^2}
\label{umix2}
\eea

\noindent
With current experimental data (\ref{expnos}) and adding the errors in
quadrature,
the limits on $|V_{\mu 1}|^2$ translate to
\begin{eqnarray}
|V_{\mu 1}|^2 &=& 0.15\pm 0.07+(0.35\pm 0.15)|V_{e3}|^2 \nonumber\\
&\pm& (0.46\pm 0.03)|V_{e3}|~.
\label{umu1val}
\end{eqnarray}
If we use the $3~\sigma$ limit of 0.05 on $|V_{e3}|^2$, we obtain a lower limit
of 0 and an upper limit of 0.35 on $|V_{\mu 1}|^2$ while the $1~\sigma$ limit
of 0.012 on $|V_{e3}|^2$ leads to $0.029\leq |V_{\mu 1}|^2\leq 0.28$.
Therefore, we can arrive at closed bound
\be
0.17\leq |V_{\mu 1}|\leq 0.52 \mbox{ at}~1\sigma~.
\label{umu1bound}
\ee
We can use this expression to set a bound on $|V_{\mu 1}|^2$ or we can invert
it to get information on $cos(\delta)$ once we have information on all four
$|V_{ij}|^2$, the exact relation being:
\begin{eqnarray}
\cos\delta& =&
 \frac{|V_{\mu1}|^2-|V_{e2}|^2(1-|V_{\mu 3}|^2) + B~|V_{e3}|^2 + C~|V_{e3}|^4}
{X}~,\nonumber\\
X &=&2|V_{e2}||V_{e3}||V_{\mu3}|\sqrt{1-|V_{e2}|^2-|V_{e3}|^2} \nonumber\\
&\times&
\sqrt{1-|V_{e3}|^2-|V_{\mu 3}|^2}~,\nonumber\\
B &=& |V_{e2}|^2-|V_{\mu3}|^2-2|V_{\mu 1}|^2+|V_{e2}|^2|V_{\mu 3}|^2,
\nonumber\\
C &=& |V_{\mu 1}|^2 + |V_{\mu 3}|^2~.\nonumber\\
\label{delbnd}
\end{eqnarray}
\noindent
Clearly, without information on $|V_{\mu 1}|^2$, we cannot say anything
on $\delta$.

\subsection{ CP-Violation and the $|V_{ij}|$'s}
The measure of CP violation is expressed, in general, through the Jarlskog
parameter defined as \cite{jarl}
\begin{eqnarray}
J = \Im m(V_{e2} V_{\mu 3} V^*_{e3} V^*_{\mu 2})
\end{eqnarray}

Once we have chosen our four independent parameters of the mixing
matrix, $J^2$ is not independent but an explicit function of these
four parameters, namely \cite{cherif}

\begin{eqnarray}
J^2&=& (1-|V_{e3}|^2-|V_{e2}|^2)|V_{e3}|^2|V_{\mu1}|^2|V_{\mu3}|^2\nonumber\\
&-&\frac{1}{4}\big(|V_{e2}|^2-|V_{\mu1}|^2+|V_{\mu1}|^2|V_{e3}|^2-|V_{\mu3}|^2|V_{e3}|^2\nonumber\\
&-& |V_{\mu3}|^2|V_{e2}|^2 \big)^2~.
\label{j1}
\end{eqnarray}
\label{Jcarre}

This is a general expression based on rephasing invariants and we can use it
to set limits on $|V_{\mu 1}|^2$: requiring positivity of $J^2$ leads to
(\ref{umix2}) with $\cos\delta$ replaced by $\pm 1$. We can also calculate
what value of $|V_{\mu 1}|^2$ will maximize $J^2$ such that
$|V_{\mu 1}|^2_{J^2=J^2_{max}}$. Differentiating $J^2$ with respect to
$|V_{\mu 1}|^2$ and retaining only first order in
$|V_{e3}|^2$, one
then recovers the first two terms of (\ref{umix2}). This is consistent since
maximizing $J^2$ requires $\sin^2\delta = 1$. One observes
here that a value of $|V_{\mu 1}|^2$ larger than
$|V_{\mu 1}|^2_{J^2=J^2_{max}}$
requires $\delta$ to be in the first or in the fourth quadrant.
This simple point does not require
particularly precise data and could already be useful information for model
builders.

In the standard basis, $J^2$ is given by \cite{jsq}
\begin{eqnarray}
J^2 &=&sin^2(\theta_{12})cos^2(\theta_{12})~
     sin^2(\theta_{23})cos^2(\theta_{23})sin^2(\theta_{13})~\nonumber\\
&\phantom{=}&cos^2(\theta_{13})~cos^2(\theta_{13}) sin^2(\delta)
\end{eqnarray}

Clearly, whether in the $V_{PMNS}$ or in our current basis, $J^2$ is the same
physical observable and we can compare the two expressions for this
parameter.
Using the relations between the $V_{ij}$ and $sin(\theta_{ij})$,
we express
$J^2$ in terms of $sin^2(\delta)$ and three of our $|V_{ij}|^2$:
\begin{eqnarray}
 J^2 &=&\frac{1}{(1-|V_{e3}|^2)^2}|V_{e2}|^2 |V_{\mu3}|^2|V_{e3}|^2( 1- |V_{e3}|^2-|V_{e2}|^2)\nonumber\\
  &\times&(1-|V_{e3}|^2-|V_{\mu3}|^2)\sin^2\delta~.
  \label{j2}
\end{eqnarray}

Using (\ref{j1}) and (\ref{j2}) we now have a relation between $\sin^2\delta$ and our
four $|V_{ij}|^2$.
Therefore, we can relate $\delta$ to our set of four $|V_{ij}|^2$ through
either $\cos\delta$ or $\sin^2\delta$. Since we want ultimately information
on $\delta$ itself and there is more information through $\cos\delta$ than
through $sin^2(\delta)$, in what follows, we will concentrate on $\cos\delta$.
We will be left with a twofold degeneracy as we will not be able
to get the sign of $\delta$.

On figure 1, we plot $J^2$ as a function of $|V_{\mu 1}|^2$ and
$|V_{e3}|^2$ for central values of $|V_{\mu 3}|^2$ (0.5) and
$|V_{e2}|^2$ (0.3). We see clearly that the vanishing of $|V_{e3}|^2$ leads
to the vanishing of $J^2$, as it must be. We also see that $J^2$ is a fairly
linear function of $|V_{e3}|^2$; this is expected since the maximum value
$|V_{e3}|^2$ can have is rather small compared to the other parameters and
an expansion to first order in $|V_{e3}|^2$ would be adequate. For numerical
purposes, we note that the largest value taken by $J^2$ on this figure is
$2.42\times 10^{-3}$; it is clearly at the largest possible value of
$|V_{e3}|^2$ and at $|V_{\mu 1}|^2\simeq 0.167$

On figure 2, we plot curves of constant $sin^2(\delta)$. When we fix
$|V_{e2}|^2$, $|V_{\mu 3}|^2$, and $|V_{e3}|^2$, if we ask what value of
$|V_{\mu 1}|^2$ will give $25\%$ of the maximum value that $J^2$ can have,
we are infact setting $sin^2(\delta) = 0.25$. The straight line in the
middle is the curve $|V_{\mu 1}|^2_{J^2=J^2_{max}}$ and is given by
\ref{umix2} with $cos(\delta) = 0$.

\section{ Constraining $\delta$ from the $|V_{ij}|^2$ or their Ratios}
In order to be able to set some bounds on $\delta$, we need at least lower
limits on $|V_{e3}|^2$ and $|V_{\mu 1}|^2$ besides the data that we have on
$|V_{e2}|^2$ and $|V_{\mu 3}|^2$; as we can see on figure 2, just an upper
limit cannot constrain $\delta$.
In what follows, we explore the full 4-dimensional parameter
space: we allow $|V_{\mu 1}|^2$ to cover the whole range allowed by unitarity,
and $|V_{e3}|^2$ varies from 0.001 to 0.03,
since 0.03 is close to the current 3-$\sigma$ limit and 0.001 makes
CP-violation extremely small at planned CP-violating experiments\cite{cpexp}.
As for $|V_{e2}|^2$ and
$|V_{\mu 3}|^2$, we will work with values that can differ by $\pm 0.1$
from their current central values of 0.3 and 0.5, respectively.

 We assume that we have four $|V_{ij}|^2$ with their experimental
uncertainties. The experimental central values of the $|V_{ij}|^2$'s
($|V_{ij}|^2_c$) lead to the central value of $\cos\delta$. In order to
estimate the error, or range that we should associate to this central value,
we use a Monte Carlo technique and we cover, for each $|V_{ij}|^2$, the space
($|V_{ij}|_c^2 - \mbox{experimental error}$) - ($|V_{ij}|_c^2 + 
\mbox{experimental error})$.
Within this 4-dimensional space, we are interested
only in the largest and smallest values that $\cos\delta$ can have;
these become the range that $\cos\delta$ covers with this particular
combination of $|V_{ij}|^2$ and their associated errors.

As our fourth $|V_{ij}|^2$, we picked $|V_{\mu 1}|^2$.
This choice is not unique and we could have chosen $|V_{\mu 2}|^2$,
or $|V_{\tau 1}|^2$, or $|V_{\tau 2}|^2$. It is straightforward to rewrite
all our equations in terms of these parameters through the following
relations:
\begin{eqnarray}
|V_{\mu 2}|^2&=& 1 - |V_{\mu 1}|^2 - |V_{\mu 3}|^2\nonumber\\
|V_{\tau 1}|^2& =& |V_{e2}|^2 + |V_{e3}|^2 - |V_{\mu 1}|^2\nonumber\\
|V_{\tau 2}|^2& =& |V_{\mu 3}|^2 + |V_{\mu 1}|^2 - |V_{e2}|^2
\label{vijik}
\end{eqnarray}

We studied all four parameters and we can say that, when they
({\it ie} $|V_{\mu 1}|^2,|V_{\mu 2}|^2,|V_{\tau 1}|^2,|V_{\tau 2}|^2$) have
the same experimental error ({\it eg} 5\%) then:
\hfil\hfil\break
\noindent$\bullet$~if the known parameters stay close to their current
central values, in general, $|V_{\mu 1}|^2$ leads to slightly better bounds
than $|V_{\tau 1}|^2$
\hfil\hfil\break
\noindent$\bullet$~if the known parameters digress substancially from their
current experimental values, then $|V_{\tau 1}|^2$ leads to slightly
better bounds than $|V_{\mu 1}|^2$
\hfil\hfil\break
\noindent$\bullet$~$|V_{\mu 2}|^2$ and $|V_{\tau 2}|^2$ are not as good as the
previous two: the limits obtained from these parameters are degraded by about a
factor of 2 when compared to those obtained from the previous two parameters;
$|V_{\mu 2}|^2$ is, in general, a little bit better than $|V_{\tau 2}|^2$
\hfil\hfil\break
\noindent$\bullet$~ things improve a bit for the last two parameters if we
give all four parameters the same absolute uncertainty instead of the same
relative uncertainty ({\it eg} we compare $|V_{\mu 1}|^2 = 0.150\pm 0.005$ and
$|V_{\tau 2}|^2 = 0.350\pm 0.005$ instead of $\pm 5\%$ for all parameters).
Even then, $|V_{\mu 1}|^2$ and $|V_{\tau 1}|^2$ remain better than
the other two parameters, but by a factor of 1.5 instead of a factor of 2.

We also take into consideration some
ratios of $|V_{ij}|^2$ as potential fourth parameter.
The first one is $\rho_1 = |V_{\tau 1}|^2/|V_{\mu 1}|^2$ and has very similar
properties to $|V_{\tau 2}|^2/|V_{\mu 2}|^2$; the second one is
$\rho_2 = |V_{\tau 2}|^2/|V_{\tau 1}|^2$ and has very similar properties to
$|V_{\mu 2}|^2/|V_{\mu 1}|^2$. We also studied
$|V_{\tau 2}|^2/|V_{\mu 1}|^2$ but it turned out to be too sensitive to
both $|V_{e3}|^2$ and $|V_{\mu 3}|^2$ to be of any use; and similarly for
$|V_{\tau 1}|^2/|V_{\mu 2}|^2$. The same can be said of any ratio that
involves $|V_{e1}|^2$. So, we
present in Tables I, and II the limits that we can set on $\cos\delta$
from measurements on $|V_{\mu 1}|^2$, $\rho_1$ and $\rho_2$. These are
representative of what can be achieved. After studying the parameter space
described above, we can say that:
\hfil\hfil\break
\noindent
$\bullet$ uncertainties of $10\%$ on the four parameters can lead to
very tight bounds on $\delta$; the uncertainty on $|V_{e3}|^2$ can be
much larger than this without affecting the bounds very much.
\hfil\hfil\break
\noindent
$\bullet$ an interesting constraint already occurs for the combination
(0.3,0.5,$P$,$|V_{e3}|^2$) with a 10\% uncertainty on the parameters:
the first two parameters ($|V_{e2}|^2$ and $|V_{\mu 3}|^2$)
are at their current central values while the third one (either
$|V_{\mu 1}|^2=0.15$, $\rho_1=1.0$ or $\rho_2=2.3$)
is close to the value that maximizes $J^2$. We find that if $|V_{e3}|^2$
turns out to be large (0.03), then $|\delta|$ has a range of 40-70 degrees
centered at about 90 degrees, depending what $P$ one uses; this range
decreases as $|V_{e3}|^2$ increases. If the uncertainties are reduced to 5\%,
the range becomes 30-60 degrees and $|V_{e3}|^2$ can be reduced to 0.01.
\hfil\hfil\break
\noindent
$\bullet$ for most combinations of $|V_{e2}|^2$ and $|V_{\mu 3}|^2$, if
$|V_{\mu 1}|^2$ turns out to be relatively small, about 0.1, then $|\delta|$
has to be between $\pi/2$ and $\pi$ for any $|V_{e3}|^2$ larger than 0.005.
\hfil\hfil\break
\noindent
$\bullet$ the limits are not very sensitive to the uncertainty on
$|V_{e3}|^2$; going from $5\%$ to $20\%$ does not change the limits by much.
If the other uncertainties are small ($2\%$), the uncertainty on
$|V_{e3}|^2$ can go up to 50$\%$ and it is still possible to get interesting
bounds on $\delta$
\hfil\hfil\break
\noindent
$\bullet$ the limits are not very sensitive to $|V_{e3}|^2$ itself and
going from 0.01 to 0.005 will not change the limits very much. When
$|V_{e3}|^2$ becomes 0.001, however, the limits are somewhat degraded
but still useful.
\hfil\hfil\break
\noindent
$\bullet$ as the value of $|V_{e3}|^2$ becomes smaller, the uncertainties
on the other parameteres must decrease in order to keep interesting bounds on
$\delta$; note that the uncertainty on $|V_{e3}|^2$ can be rather large
without affecting much the bounds
\hfil\hfil\break
\noindent
$\bullet$ in general, we find that the best parameter is $\rho_1$, the
second best is $\rho_2$ and the third best is $|V_{\mu 1}|^2$.
\hfil\hfil\break
\hfil\hfil\break
\noindent
The previous analysis is very general and describes which parameters must be
measured and with what accuracy in order to set limits on $\delta$.
We now turn to some potential processes that could give us the fourth
information that we need in order to be able to set limits on $\delta$.

\section{Ultra High Energy Cosmic Rays}
Ultra high energy cosmic rays and their detectors have attracted a fair
amount of attention in recent years.\cite{jb,nutel,katrin,irina} In
particular, the use of UHCRs with the aim of observing leptonic CP-violation
has been considered in \cite{farza}
An interesting consequence for the UHCR neutrino spectra is due to maximal
atmospheric mixings. Following maximal mixing, or $\tan\theta_A \approx 1$
leads to a unique prediction for UHCR neutrino fluxes. It was shown
that UHCR neutrinos (which are expected to be sourced by cosmic objects
such as AGNs) when measured by ground based detectors, the expected flavor
ratio $\phi_e:\phi_\mu:\phi_\tau=1:1:1$
\cite{jukka}. This value is also known as the standard flavor ratio and
by itself constitute
an independent confirmation of the neutrino mixing data from atmopheric
sources. Since we propose to extract the Dirac
CP phase present in the conventional PMNS mixing matrix,
our discussion will be restricted to
Dirac neutrinos. Before proceeding with the analysis, let us briefly allude
to the significance of this proposal. It is
well known that CP violation (due to oscillation) in the leptonic mixings
(even if larger than the
quark secotr) will nonetheless remain a hard problem to resolve\cite{cpexp}.
This arises from
the strong constraints which reactor neutrinos set on the mixing element
$V_{e3}$ \cite{chooz}. As a consequence of this, even if the CP phase is
large, the smallness of $V_{e3}$ leads to phase being
insensitive to any CP violating measurements.

\subsection{The basic formalism}
In the same fashion as quarks mix, massive neutrinos mix
(and also oscillate) between two eigenbas2s,
the flavor $(\nu_\alpha)$ and mass
eigenbases $(\nu_i)$. The two bases are related by a unitary matrix $V$
such that $\nu_\alpha = V_{\alpha i} \nu_i$. Here we assume the summation
over the mass eigenstates.
Corresponding to a particular mass eigensates is a mass value,
$m_i$ which in the limit of small mixings determines the mixing angles.
Thus, in the limit of small mixings, $\theta_S$ mixes eigenstates $\nu_1$
and $\nu_2$, while $\theta_A$ mixes  eigenstates $\nu_2$ and $\nu_3$
and $\theta_R$ mixes the eigenstates $\nu_1$ and $\nu_3$.

In the context of UHCR neutrinos which travel astronomical distances,
the coherence between the various mass eigenstates is averaged out.
In other words, after production these neutrinos essentially travel as
individual mass eigenstate. At the point of detection, they are in the flavor
states. Therefore, the detection probability in a given flavor eigenstate
is
\bea
\phi_e &=& 1 + 2x(2c_A^2-1)~;~x=(s_Sc_S)^2~,\nonumber\\
\phi_\mu &=& 2xc_A^2 + 2(c_A^4(1-2x) +s_A^4)~,\nonumber\\
\phi_\tau &=& s_{A}^2 + 2xs_A^2(1-c_A^2)~,
\label{flux}
\eea
where $s$ and $c$ denote sine and cosine, respectively. From the above
expressions and the experimental fact $\theta_A\simeq\pi/4$,
it follows that all neutrino flavors
must be detected with the same weight factor. In deriving this result,
we have disregarded the mixing corresponding to reactor experiments, which is
consistent with zero \cite{chooz}.
This result is also independent of the solar angle $\theta_S$.

\subsection{Neutrino Fluxes and $|V_{ij}|^2$}
Let us revise in more detail the arguments from the previous section.
Consider the probability of oscillating from flavor $\alpha$ to $\beta$
\begin{eqnarray}
P_{\alpha \beta} &=&\delta_{\alpha\beta} - 4 \sum_{i> j=1}^3
Re[K_{\alpha\beta,ij}]
\sin^2(\Delta_{ij})\nonumber\\
&+&4 \sum_{i> j=1}^3 Im[K_{\alpha\beta,ij}]~sin(\Delta_{ij})
\cos(\Delta_{ij})~,\nonumber\\
K_{\alpha\beta,ij} &= &V_{\alpha i}V_{\beta i}^*V_{\alpha j}^*V_{\beta j}~;~
\Delta_{ij} = \frac{m_i^2 - m_j^2}{L/4E}~.
\label{prob}
\end{eqnarray}

We can express the real and imaginary parts of $K_{\alpha\beta,ij}$
in terms of moduli as follows:
\begin{eqnarray}
2 Re[K_{\alpha\beta,ij}]&=& |V_{\alpha i}|^2|V_{\beta j}|^2 +
|V_{\beta i}|^2|V_{\alpha j}|^2\nonumber\\
& -& \sum_{\gamma,k} c_{\alpha\beta\gamma}c_{ijk}|V_{\gamma k}|^2
\end{eqnarray}
\noindent
and
\begin{eqnarray}
Im[K_{\alpha\beta,ij}] = J~\varepsilon_{\alpha\beta}~\varepsilon_{ij}
\end{eqnarray}
\noindent
where we have defined:
\bea
c_{ijk} &=& 1~{\rm if}~i\ne j,~j\ne k,~k\ne i\nonumber\\
        &=& 0~{\rm otherwise}
\eea
\noindent
and $\varepsilon$ as the antisymmetric tensor
\bea
\varepsilon = \pmatrix{\phantom{-}0 & -1 & \phantom{-}1 \cr
\phantom{-}1 & \phantom{-}0 & -1 \cr
-1 & \phantom{-}1  & \phantom{-}0 \cr}
\eea

We are interested here only in the probabilities for
$\nu_\mu\to\nu_\tau$, $\nu_\mu\to\nu_e$ and $\nu_\tau\to\nu_e$. After
combining two terms ($\Delta_{13} = \Delta_{23}$ to high accuracy),
expressing the third term as a combination of $|V_{ij}|^2$ and averaging over
the oscillating terms, we obtain
\begin{eqnarray}
P_{\mu\tau} &=&2 |V_{\mu 3}|^2 |V_{\tau 3}|^2
-(|V_{e1}|^2|V_{e2}|^2 - |V_{\mu 1}|^2|V_{\mu 2}|^2\nonumber\\
& - & |V_{\tau 1}|^2|V_{\tau 2}|^2) ~,\nonumber\\
P_{e\tau} &=& 2 |V_{e 3}|^2 |V_{\tau 3}|^2
-(|V_{\mu 1}|^2|V_{\mu 2}|^2 - |V_{e1}|^2|V_{e2}|^2 \nonumber\\
&-& |V_{\tau 1}|^2|V_{\tau 2}|^2) ~, \nonumber\\
P_{e\mu} &=&2 |V_{\mu 3}|^2 |V_{e3}|^2
-(|V_{\tau 1}|^2|V_{\tau 2}|^2 - |V_{\mu 1}|^2|V_{\mu 2}|^2 \nonumber\\
&-& |V_{e1}|^2|V_{e2}|^2) ~.
\end{eqnarray}

One then uses the relations
\begin{eqnarray}
|V_{e1}|^2 &=& 1 -|V_{e2}|^2- |V_{e3}|^2 ~,\nonumber\\
|V_{\mu 2}|^2 &= & 1 - |V_{\mu 1}|^2- |V_{\mu 3}|^2~,\nonumber\\
|V_{\tau 1}|^2 &= & |V_{e2}|^2 + |V_{e3}|^2- |V_{\mu 1}|^2~,\nonumber\\
|V_{\tau 2}|^2& = & |V_{\mu 1}|^2 + |V_{\mu 3}|^2- |V_{e2}|^2~.\nonumber\\
\end{eqnarray}
to express these probabilities in terms of our four $|V_{ij}|^2$.

Note that when averaging over the oscillating terms, the third term of
\ref{prob} vanishes; therefore, one will not be able to observe directly
CP-violation with UHCR's.

If we assume that the main production mode of these UHCR is \cite{irin}
\begin{eqnarray}
p + \gamma \to \pi^\pm + X ~,
\end{eqnarray}
with subsequent decay of the $\pi^\pm$ into muons, electrons, and neutrinos,
we conclude that there will be two $\nu_\mu$ (or $\bar\nu_\mu$) for every
$\nu_e$ (or $\bar\nu_e$) and virtually no $\nu_\tau$ (or $\bar\nu_\tau$).
The initial fluxes are then $\phi_e^0$, $\phi_\mu^0 = 2\phi_e^0$, and
$\phi_\tau^0 = 0$. In order to calculate a given neutrino flux that reaches the
earth, we take into account the probability that this neutrino will oscillate
into other types of neutrinos and the probabilities that other neutrinos will
oscillate into this type of neutrino. Up to a common geometrical
factor, the  observed terrestrial fluxes are:
\begin{eqnarray}
\phi_e^t &=& \phi_e^0~(1 + P_{e\mu} - P_{e\tau})~,\nonumber\\
\phi_\mu^t &=& \phi_e^0~(2 - P_{e\mu} - 2P_{\mu\tau})~,\nonumber\\
\phi_\tau^t& =& \phi_e^0~(2P_{\mu\tau} + P_{e\tau})~.
\end{eqnarray}
\noindent
In terms of matrix elements, we get
\begin{eqnarray}
\phi_e^t &=& \phi_e^0~\big(
1- 2|V_{e2}|^2(|V_{\mu 3}|^2-|V_{e2}|^2)\nonumber\\&+&
2|V_{\mu 1}|^2(1-2|V_{e 2}|^2)\nonumber\\
&-&2|V_{e3}|^2(1 + |V_{\mu 1}|^2 -  |V_{e2}|^2 - |V_{\mu 3}|^2)
+2 |V_{e3}|^4\big)~,\nonumber\\
\phi_\mu^t &=& \phi_e^0~\big(
2 +(1-|V_{\mu 3}|^2)(|V_{e2}|^2 -4|V_{\mu 3}|^2)\nonumber\\
&-&|V_{\mu 1}|^2(3 + 2|V_{e2}|^2-4|V_{\mu 3}|^2 - 4|V_{\mu 1}|^2)\nonumber\\
&+&|V_{e3}|^2(|V_{\mu 3}|^2 - |V_{\mu 1}|^2)\big)~,\nonumber\\
\phi_\tau^t &=& \phi_e^0~\big(4|V_{\mu 3}|^2(1-|V_{\mu 3}|^2) -
|V_{e2}|^2(1-|V_{e2}|^2) \nonumber\\ &+&
3|V_{e2}|^2(|V_{\mu 3}|^2)- |V_{e2}|^2)\nonumber\\
&+&|V_{e3}|^2(2 + 3|V_{\mu 1}|^2 - 3|V_{\mu 3}|^2 - 2|V_{e2}|^2 - 2|V_{e3}|^2)
\nonumber\\
&+&|V_{\mu 1}|^2(1 - 4|V_{\mu 1}|^2 - 4|V_{\mu 3}|^2 + 6|V_{e2}|^2)\big)
\end{eqnarray}

Clearly, we don't know the original neutrino fluxes at the source. Therefore,
it is more meaningful to try to measure ratios of neutrino fluxes.
We will consider the ratios $R_{e\mu} = \phi_e^t /\phi_\mu^t$ and
$R_{e\tau} = \phi_e^t /\phi_\tau^t$ and calculate how they vary when we cover
the parameter space available. Our results are summarized in Table III. One
can see that:
\hfil\hfil\break
\noindent$\bullet$ by numerical accident, $|V_{\mu 3}|^2$ close to 0.5 has a
very strong influence on $R_{e\mu}$ and $R_{e\tau}$ and tends to keep them
close to 1
\hfil\hfil\break
\noindent$\bullet$ if $|V_{\mu 3}|^2$ stays close 0.5 to within 10$\%$,
$R_{e\mu}$ and $R_{e\tau}$ remain close to 1, for any value of $|V_{e2}|^2$ and $|V_{e3}^2|$ that are within the experimentally allowed ranges
\hfil\hfil\break
\noindent$\bullet$ when one varies $|V_{\mu 3}|^2$ by 20$\%$, either up or
down, it is possible for $R_{e\mu}$ and
$R_{e\tau}$ to differ from 1 by as much as 20$\%$. Unfortunately, these
large discrepancies require values of $|V_{\mu 1}|^2$ such that
$\delta\simeq 0~{\rm or}~\pi$, which means very little CP-violation
\hfil\hfil\break
\noindent$\bullet$ $|V_{e3}|^2$ has very little impact and varying it from
0.01 to 0.005 will change $R_{e\mu}$ and $R_{e\tau}$,
typically by 1$\%$ or so.
\hfil\hfil\break
\hfil\hfil\break
\noindent
Therefore, we can say that even within the standard production mechanism
for the UHCR, deviations of $R_{e\mu}$ and $R_{e\tau}$ by
20$\%$ from their expectation value of 1 are possible with the current
experimental data but require in general pushing these data to their limits
and CP-violation to be vanishingly small.

\subsection{Neutrino Fluxes and $\delta$}
In order to study the bounds one could set on $\delta$ if one were to measure
$R_{e\mu}$ with a given accuracy, we express $|V_{\mu 1}|^2$ in
terms of $R_{e\mu}$ and use that result directly into our expression of
$cos(\delta)$~(\ref{delbnd}). Again, we use a Monte Carlo technique to scan
the allowed parameter space.
As a technical detail, we note that we have a quadratic
equation to solve with two possible values for $|V_{\mu 1}|^2$ but only one
value respects unitarity and leads to a positive value of $J^2$; events that
do not respect unitarity are excluded from our Monte Carlo.
In Table IV, we summarise our results. We see that:
\hfil\hfil\break
\noindent
$\bullet$ the most important
parameters are $|V_{\mu 3}|^2$ and $R_{e\mu}$, where the uncertainties must
be rather small, while $|V_{e2}|^2$ and $|V_{e3}|^2$ can tolerate much larger
errors without degrading the limits much
\hfil\hfil\break
\noindent
$\bullet$ in general, a precision of two
percent on all 4 parameters leads to tight constraints on $\delta$;
this can be relaxed to $10\%$ for $|V_{e2}|^2$ and up to $25\%$ for
$|V_{e3}|^2$ without loosing much on the bounds on $\delta$
\hfil\hfil\break
\noindent
$\bullet$ large or small values of $R_{e\mu}$ (1.13 or 0.84, for example)
require higher precision in the four parameters in order to keep
$cos(\delta)$ away from +1 or -1; the precision required on $|V_{e2}|^2$ and
$|V_{e3}|^2$ is still lower than that on the other two parameters. Recall that
$cos(\delta)=\pm 1$ means no CP-violation.
\hfil\hfil\break
\noindent
$\bullet$ as $|V_{e3}|^2$ gets smaller, a higher precision on the parameters
is necessary in order to keep the same tight range on $\delta$
\hfil\hfil\break
\noindent
$\bullet$ with some combinations of parameters, it is possible to exclude
$\delta = \pi{\rm ~or~}0$ with a precision of a few percent on the parameters
\hfil\hfil\break
\noindent
$\bullet$ an uncertainty of $5\%$ on $R_{e\mu}$ can limit $\delta$ to
a 90-degree range or smaller with certain combination of parameters
\hfil\hfil\break
\hfil\hfil\break
The precision required in order to get interesting constraints on $\delta$ is
daunting but this was expected since we saw that $|V_{\mu 1}|^2$ was the least
sensitive of our three parameters to set limits on $\delta$.

\section{ New physics and $\delta$}
This general behaviour of cosmic neutrinos fluxes has been noted before
\cite{jukka} and
comes directly from the fact that $|V_{\mu 3}|^2$ is close to 0.5. As
$|V_{\mu 3}|^2$ goes away from 0.5, (current data allow a deviation of up to
$20\%$) the ratios can differ substantially from 1:1:1, as we just saw.
This brings the question of how one would interpret ratios that would differ
much more than 8$\%$, say 30$\%$ or more. It would be difficult to
explain these anomalies within the SM and such discrepancies might
suggest some new physics.
This new physics could bring its own CP-violation, which could be
{\it \`a la Dirac}; we would then deal with an effective CP-phase such that
$\delta_{effective} = \delta + \delta_{new}$.

We will consider new physics
beyond massive neutrinos and for illustration purposes,
we choose non-radiative neutrino decays.
\subsection{Non-radiative neutrino decays and $\delta{eff}$}
The notion of non-radiative neutrino decay as a source of deviations from
the standard ratio was first discussed in \cite{beacom} and we refer to this
paper for details. Essentially, in this model, the observable flavor ratio is
modified to be
\be
\phi_e:\phi_\mu:\phi_\tau=|V_{e1}|^2: |V_{\mu 1}|^2:|V_{\tau 1}|^2 ~,
\label{frdecay}
\ee
where we assume normal mass hierarchy.

Contrary to the SM flux ratios, the ratio $R_{\tau\mu} =
\phi_\tau/\phi_\mu$ is a direct measure of \be \rho_1 =
\frac{|V_{\tau 1}|^2}{|V_{\mu1}|^2}~. \label{rhodefn} \ee which
proved to be the best variable to set constraints on $\delta$.
Table V shows what kind of limits one could obtain with such a
scenario. We note that: \hfil\hfil\break \noindent $\bullet$
uncertainties in the $5\%$ - $10\%$ range can lead to interesting
constraints on $\delta_{eff.}$ \hfil\hfil\break \noindent
$\bullet$ the limits are insensitive to the uncertainty on
$|V_{e3}|^2$ \hfil\hfil\break \noindent $\bullet$ in this model,
flux ratios of up to 3 are allowed with current data. This limit
depends on $|V_{e3}|^2$ and as $|V_{e3}|^2$ decreases, so does the
upper limit of the ratio.

As we just saw, if $R_{e\mu}$ or $R_{e\tau}$ is close to 1, there
is no need for new physics and a precision of $1\%$ - $2\%$ is
required to set bounds on $\delta$. Such a precision appears
extremely difficult at planned detectors such as IceCube, Antares,
Nestor, Anita, or Baikal \cite{mocioiu}. An optimistic view would
be the detection of a few tens of events per year at IceCube, for
example. If we assume that all these detectors will have
comparable performance after a few years of running, then we could
hope for about 100 events per year. This will not be sufficient to
reach a precision of $1\%$ in a reasonable time frame but a $10\%$
precision appears within reach. Therefore, the better scenario,
when considering the experimental precision, is to have very
different neutrino flux ratios; in that case, a $10\%$ precision
leads to interesting constraints on $\delta_{effective}$.


\section{Conclusions}

In this paper we have shown how it is possible to get information about
CP-violation in the leptonic sector without performing any direct experiment
on CP-violation. In the usual parametrisation of the leptonic mixing, the
mixings are described by three mixing angles and the fourth parameter is the
CP-violating phase, $\delta$.
All the information about mixing and CP-violation can
also be described completely by four mixing elements, four $|V_{ij}|^2$.
We have shown how one can extract information on $sin^2(\delta)$ and
$cos(\delta)$ from four $|V_{ij}|^2$ or three $|V_{ij}|^2$ and a ratio of
two $|V_{ij}|^2$. We used two such ratios:
$\rho_1 = |V_{\tau 1}|^2/|V_{\mu 1}|^2$ and
$\rho_2 = |V_{\tau 2}|^2/|V_{\tau 1}|^2$.


We find that if one uses $|V_{e2}|^2$, $|V_{\mu 3}|^2$ and
$|V_{e3}|^2$, then it is better to use $|V_{\mu 1}|^2$ or
$|V_{\tau 1}|^2$ as the fourth independent $|V_{ij}|^2$. When the
first two parameters remain close to their current central
experimental values, $|V_{\mu 1}|^2$ is slightly better than
$|V_{\tau 1}|^2$ while $|V_{\tau 1}|^2$ is slightly better when
the first two parameters digress from their current central
experimental values.

 As a first constraint that does not require particularly
precise data, we find that if $|V_{\mu 1}|^2$ were larger than
$|V_{\mu 1}|^2_{J^2 = J^2_{max}}$, then $\delta$ would have to be
in the first or fourth quadrant.

In general, when the four parameters have uncertainties of 10\%, it is
possible to get interesting constraints on $\delta$ on some of the parameter
space. For example, assuming 10\% errors,
if $|V_{e2}|^2$ and $|V_{\mu 3}|^2$ keep their current experimental central
values and $|V_{\mu 1}|^2$ turns out to be about 0.15, then $|\delta|$
has a 70 degree range centered at about 90 degrees if $|V_{e3}|^2$ turns out
to be 0.03; this range decreases to about 40 degrees if one measures
$\rho_1$ and it decreases as $|V_{e3}^2$ increases. If the uncertainties are
reduced to 5\%, then the ranges on $\delta$ become 30-60 degrees and can
tolerate a value of 0.01 for $|V_{e3}|^2$. We also found
that, for most combinations of $|V_{e2}|^2$ and $|V_{\mu 3}|^2$, a small
value of $|V_{\mu 1}|^2$ (0.1) implies that $|\delta|$ is between $\pi/2$ and
$\pi$. It turns out that $cos(\delta)$ is not particularly sensitive to
$|V_{e3}|^2$ over most of the parameter space. This implies that once the
errors are reduced to the 2\% level, even if the error on $|V_{e3}|^2$
stayed at 10\%, it would  not degrade the constraints on $\delta$ by
much. Of the three variables investigated here, the best one turned out to
be $\rho_1$.

In order to get the fourth piece of information needed to set bounds on
$\delta$, we explored the use of ultra high energy cosmic neutrinos.
Generally, due a numerical accident stemming mostly from
$|V_{\mu 3}|^2\simeq 0.5$,
the cosmic neutrino fluxes are very similar once they
reach the earth. However,
current data allow cosmic neutrino fluxes to differ from their
central, expected value of 1 by up to $20\%$: this requires pushing the
parameters to the acceptable limits and implies that CP-violation in the
leptonic sector is vanishingly small. Unfortunately, the uncertainties on the
data would have to be at the few percent level on the cosmic neutrino fluxes
and on $|V_{e2}|^2$ in order to set interesting limits on $\delta$. Finally,
we studied a non-radiative decay model of neutrinos that could lead to flux
ratios very different from 1. Indeed, in this model, the current data allow
for neutrino fluxes to differ by a factor of up to three. Fortunately, in
this model, one ratio of neutrino fluxes is $\rho_1$ and leads to rather
interesting constraints on $\delta_{eff.}$ with experimental uncertainties
of $10\%$ on ratios of cosmic neutrino fluxes.

It appears that $1\%$ - $2\%$ precision on extremely high energy
neutrino fluxes will be out of reach for the planned detectors but
that $10\%$ precision will be within reach. It would then be
difficult to constrain $\delta$ within the SM context using these
cosmic rays and on would have to rely on other means to measure
$|V_{\mu 1}|^2$ or $|V_{\tau 1}|^2$ or $\rho_1$ or $\rho_2$. On
the other hand, if the neutrino fluxes are very different, then
one is likely outside the SM  and a $10\%$ precision on the
neutrino flux ratios is sufficient in some models to set
interesting bounds on $\delta_{effective}$.

{\bf Acknowledgements}: We wish to thank Dr. Irina Mocioiu for
useful and helpful comments. G. C. wants to thank the Watson's for
technical support. This work is funded by NSERC (Canada) and by
the Fonds de Recherche sur la Nature et les Technologies du
Qu\'ebec.

\vfil\vfil\eject
\noindent
{\bf Figure Captions}
\hfil\hfil\break
Figure 1
\hfil\hfil\break
\noindent
$J^2$ as a function of $|V_{e3}|^2$ and $|V_{\mu 1}|^2$ for the current
central values of $|V_{e2}|^2$ (0.3) and $|V_{\mu 3}|^2$ (0.5)
\hfil\hfil\break
\noindent
Figure 2
\hfil\hfil\break
\noindent
Different values of $sin^2(\delta)$ in the $|V_{e3}|^2 - |V_{\mu 1}|^2$ plane.
>From the outside, inward, we plot $sin^2(\delta) = 0, 0.1, 0.25, 0.5, 0.75,
0.95, 1$. The line in the middle corresponds to $sin^2(\delta) = 1$ and is
given by eq. 2.7 with $cos(\delta) = 0$; the most outward curves correspond to
$cos(\delta) = \pm 1$.

\begin{thebibliography}{99}
\bibitem{ffoot}
We could also use supernova neutrino data as additional observable,
but, given the possibility of detecting UHCR neutrinos more frequently than a
supernova burst, we shall not consider this option.
\end{thebibliography}
\end{document}